\documentclass{article}
\usepackage{PRIMEarxiv}
\usepackage{amsmath,amssymb,amsfonts}
\usepackage{algorithm}
\usepackage{algpseudocode}
\usepackage{graphicx}
\usepackage{textcomp}
\usepackage{cite}
\usepackage{subcaption}

\usepackage[T1]{fontenc}

\title{Learned Regularization for Microwave Tomography}
\author{
  Bowen Tong, Hao Chen, Shaorui Guo, Dong Liu\\
  University of Science and Technology of China, Hefei, China \\
  \texttt{\{bowentong, hchern, shaorui\_guo\}@mail.ustc.edu.cn} \\
  \texttt{dong.liu@outlook.com} \\
}

\begin{document}
\maketitle

\begin{abstract}

Microwave Tomography (MWT) aims to reconstruct the dielectric properties of tissues from measured scattered electromagnetic fields. This inverse problem is highly nonlinear and ill-posed, posing significant challenges for conventional optimization-based methods, which, despite being grounded in physical models, often fail to recover fine structural details. Recent deep learning strategies, including end-to-end and post-processing networks, have improved reconstruction quality but typically require large paired training datasets and may struggle to generalize. To overcome these limitations, we propose a physics-informed hybrid framework that integrates diffusion models as learned regularization within a data-consistency-driven variational scheme. Specifically, we introduce Single-Step Diffusion Regularization (SSD-Reg), a novel approach that embeds diffusion priors into the iterative reconstruction process, enabling the recovery of complex anatomical structures without the need for paired data. SSD-Reg maintains fidelity to both the governing physics and learned structural distributions, improving accuracy, stability, and robustness. Extensive experiments demonstrate that SSD-Reg, implemented as a Plug-and-Play (PnP) module, provides a flexible and effective solution for tackling the ill-posedness inherent in functional image reconstruction.

\end{abstract}

\section{Introduction}
\label{sec:introduction}

Microwave Tomography (MWT) is an emerging functional imaging modality that estimates the spatial distribution of dielectric properties in biological tissues by measuring the scattering of microwave signals.
In a typical MWT setup, multiple measurement groups are utilized, each consisting of a single transmitter emitting microwave signals and an array of receivers capturing the resulting scattered fields.
These measurements are processed to solve a nonlinear, ill-posed inverse problem, enabling reconstruction of the dielectric property distribution within the tissue.
As a non-ionizing, non-invasive, and relatively low-cost technique, MWT offers several advantages over conventional imaging methods.
Its sensitivity to dielectric contrast, particularly between malignant and healthy tissues, enables tumor detection without the need for radioactive tracers as required in modalities like Positron Emission Tomography (PET). These characteristics position MWT as a promising tool for various clinical applications, including early-stage breast cancer screening \cite{kwon2016mwi_brst_cnsr, oloumi2019mwi_3d_breast}, stroke localization \cite{guo2022stroke}, and skin cancer diagnosis \cite{mirbeik2018mwi_skin_cancer}.
\begin{figure*}
  \centering
  \includegraphics[width=1.\textwidth]{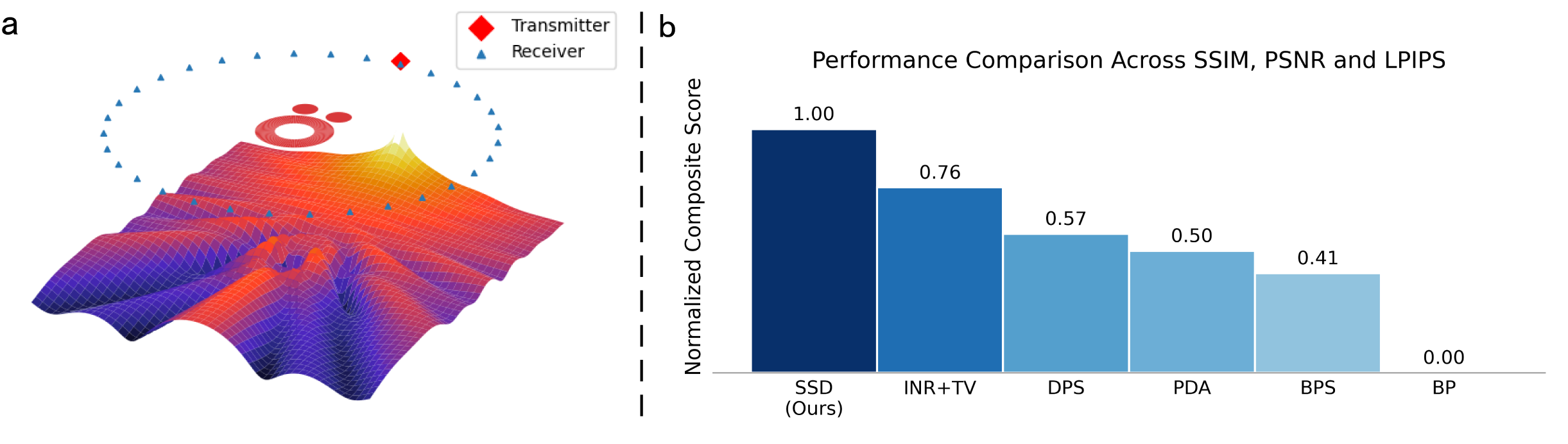}
  \caption{Schematic of MWT and performance comparison of reconstruction methods. (a) Illustration of the MWT setup, showing transmitter/receiver configuration and scattering field intensity from the Austria test case. (b) Quantitative comparison of the proposed method against representative reconstruction approaches on a simulated dataset, evaluated using SSIM, PSNR, and LPIPS metrics, as defined in Sec.~\ref{sec:metrics}}
  \label{fig:title_fig}
\end{figure*}

Despite its clinical potential, MWT remains a mathematically and computationally challenging imaging modality. The fundamental difficulties arise from the nature of the electromagnetic inverse scattering problem (EISP) \cite{xd_chen2018isp, xdchen2016isp}.
First, due to the nonlinear nature of wave propagation governed by Maxwell’s equations, even small perturbations in permittivity can lead to significant and spatially complex changes in the scattered field, particularly in the presence of multiple scattering \cite{liu2017multi_scattering}.
Second, the EISP is severely ill-posed, leading to non-unique solutions and sensitivity to measurement noise or system imperfections, which often cause unstable reconstructions, convergence issues, and image artifacts in practical settings.

Meanwhile, conventional MWT reconstruction techniques often linearize the nonlinear inverse scattering problem, limiting their ability to resolve fine structural details \cite{gao2015dbim}. Alternatively, contrast source inversion (CSI) methods \cite{van2001csi_1, sun2017csi_2} enhance reconstruction quality but weaken the physical constraints of the forward problem by avoiding the Jacobian matrix, thereby increasing demands on initial solutions and regularization. Traditional mathematical regularization approaches, such as Total Variation (TV) \cite{gonzalez2017tv}, requiring manual tuning of weighting parameters. This process becomes increasingly complex when incorporating multiple priors \cite{gholami2022tv_smth_auto}.

Recent advances in MWT reconstruction have leveraged deep learning to approximate the complex nonlinear mapping between measurement data and permittivity distributions, either through direct end-to-end models or as post-processing steps \cite{wei2018bps, xu2020deep}. While these methods demonstrate promising results under specific conditions, they often suffer from limited generalization when faced with variations in dielectric properties, sensor configurations (e.g., transmitter/receiver placements, operating frequencies), or reconstruction domain geometries \cite{sanghvi2019embedding, song2021pgan}.
To address these limitations, physics-informed neural networks, which incorporate partial differential equation constraints during training, offer a more interpretable alternative \cite{chen2024nsno, guo2021phys_isp}. However, these approaches still require large paired datasets for supervised learning.

In contrast to supervised deep learning approaches, recent research has explored implicit neural representations (INRs) to incorporate priors directly into the imaging process without relying on paired data, showing promising results in related inverse problems \cite{wang2023inr_eit, luo2024inr_eisp}. Alternatively, neural networks have also been employed as explicit regularizers within optimization frameworks to guide reconstruction \cite{2017red}. More recently, Diffusion Models (DMs) have emerged as a powerful generative framework for solving inverse problems (IPs), owing to their ability to produce high-quality and diverse samples \cite{liu2025diff_ldct, gungor2023dp_mri, gong2024pet}. Unlike traditional deep learning methods, diffusion models incorporate priors learned from large-scale image distributions, enabling enhanced reconstruction with fine-grained structural detail without requiring paired training data. While DMs have demonstrated impressive performance in tasks such as image deblurring, inpainting, and super-resolution \cite{chung2023dps, reddiff, wang2024dmplug}, their application to MWT remains largely unexplored. The severe ill-posedness and nonlinearity of the EISP pose unique challenges, often destabilizing DM-based reconstruction pipelines.

To address the challenges of ill-posedness and computational complexity in MWT, we propose a hybrid reconstruction framework that integrates accurate physical modeling with generative priors. Our approach employs a Fr\'{e}chet-differentiable forward model to enable stable and efficient optimization through precise Jacobian matrices. Concurrently, we incorporate learned priors via a single-step diffusion process, leveraging the generative power of DMs without requiring paired training data.
This prior is incorporated through a lightweight PnP regularization framework, which enhances reconstruction fidelity with reduced computational cost while guiding solutions toward anatomically plausible outcomes.

The main contributions of this work are as follows:
\begin{itemize}
    \item {\bf Hybrid Physics-Generative Framework}: We develop a reconstruction pipeline that integrates physics-guided optimization with diffusion-based generative regularization, improving both stability and accuracy in MWT.
    \item {\bf Efficient Single-Step Diffusion Regularization}: We propose SSD-Reg, a novel one-step diffusion regularization scheme that enhances reconstruction quality without the need for paired datasets.
    \item {\bf Unsupervised Diffusion Priors for MWT}: To the best of our knowledge, this work presents the first successful application of unsupervised DMs in MWT, offering a scalable and data-efficient alternative to traditional deep learning approaches.
\end{itemize}

The remainder of this paper is organized as follows: Sec.~\ref{sec:related_work} reviews related work on MWT reconstruction and diffusion-based inverse problem solving. Sec.~\ref{sec:physics} details the physical modeling of MWT and the Fr\'{e}chet-differentiable forward operator. Sec.~\ref{sec:method} presents our SSD framework and implementation. Sec.~\ref{sec:exp} provides comprehensive experimental evaluations, including comparisons with state-of-the-art (SOTA) methods on simulated and real-world data. Sec.~\ref{sec:discussions} evaluates the noise robustness, ablation studies, and discussions on its applicability to high-contrast scenarios and future research directions. Finally, Sec.~\ref{sec:conclusion} concludes the paper.


\section{Related Work}
\label{sec:related_work}
In this section, we review related works on MWT (Sec.~\ref{sec:isp}) and  Diffusion Prior for inverse problems (Sec.~\ref{sec:diffusion_isp}).

\subsection{Unsupervised Learning for EISP}
\label{sec:isp}
To address the limitations of traditional and supervised learning-based methods, unsupervised iterative reconstruction frameworks have recently gained prominence. For instance, INRs have been utilized to model permittivity and contrast sources, enhanced by positional encoding and random sampling to improve representation capacity. When combined with TV regularization, this approach has achieved SOTA performance \cite{luo2024inr_eisp}.

However, splitting the optimization loss into data and state components weakens the enforcement of the forward model constraints, often leading to unstable convergence. Moreover, the high parameter count of INRs significantly slows down optimization. Crucially, the initial contrast source is typically unknown, making it difficult to obtain a good initialization. In high-contrast scenarios, where multiple scattering is pronounced, the reconstruction becomes highly sensitive to the initial guess, frequently resulting in convergence failure. This limits the clinical applicability of such methods, especially in cases involving tumors or other high-contrast anomalies.

Unlike conventional methods, the proposed approach leverages accurate Jacobian matrices from a Fr\'{e}chet-differentiable scattering model to explicitly minimize the data-consistency (DC) loss, ensuring stable reconstructions and rapid convergence.

\subsection{Diffusion Models for Inverse Problems}
\label{sec:diffusion_isp}
DMs are generative frameworks that transform samples from a simple prior distribution (e.g., Gaussian) into a complex target distribution by reversing a gradual noising process. The forward process perturbs a clean sample $\mathbf{x}_0 \sim p_{\text{data}}$ over time steps $t = 0, 1, \ldots, T$ using a Variance Preserving Stochastic Differential Equation (VP-SDE), adding Gaussian noise according to a linear noise schedule $\beta_t = \beta_{\min} + (\beta_{\max} - \beta_{\min}) \frac{t}{T}$:
\begin{equation}
    \mathbf{x}_t = \sqrt{1 - \sigma_t^2} \mathbf{x}_0 + \sigma_t \mathbf{z}, \quad \mathbf{z} \sim \mathcal{N}(0, \mathbf{I}),
    \label{eq:x_t}
\end{equation}
where $\sigma_t = 1 - e^{-\int_0^t \beta(v) dv}$ .
The model is trained to approximate the score function $\nabla_{\mathbf{x}_t} \log p(\mathbf{x}_t)$ at various noise levels using a denoising score matching loss \cite{song2021scorebased}. Once trained, the DM can reverse the noising process via iterative sampling, progressively refining a noisy sample $\mathbf{x}_T \sim \mathcal{N}(0, \mathbf{I})$ toward a clean sample $\mathbf{x}_0$.

In the context of inverse problems, the goal is to recover an unknown signal $\mathbf{x}$ from measurements $\mathbf{y}$ governed by a forward model $\mathbf{y} = \mathcal{F}(\mathbf{x}) + \zeta$, where $\mathcal{F}$ is a known operator and $\zeta$ is noise. When $\mathcal{F}$ is linear, it can be represented as a matrix. DMs can be adapted for such tasks using either supervised or unsupervised approaches. Supervised methods train the DM using paired datasets of measurements and ground truth (GT) images \cite{bi2024diff_3d_isp}, but generalization is often limited by the training distribution.

Unsupervised DM approaches offer flexibility and improved generalization by decoupling the training of the generative prior on the signal $\mathbf{x}$ from the inverse problem, incorporating the forward model during sampling \cite{song2022inverse_score_dm, chung2023dps, reddiff, song2023pgdm}. Methods like DPS \cite{chung2023dps} use gradient corrections from the data-fidelity term to guide reverse sampling, while others integrate DMs into the forward problem to strengthen manifold constraints \cite{wang2024dmplug}.
However, reconstruction quality remains highly dependent on the quality of the learned prior, and fixed sampling schemes can be unstable when dealing with nonlinear, ill-posed inverse problems.

To overcome these issues, we adopt a single-step, inference-only framework inspired by Score Distillation Sampling (SDS) loss \cite{dreamfusion} and RED-diff method \cite{reddiff}. Instead of sampling from the reverse process, we treat the DM as a learned regularizer within a variational reconstruction scheme.
This approach avoids costly backpropagation through the entire DM and stabilizes optimization while preserving the expressiveness of the learned prior. Compared to posterior sampling, our integration achieves faster, more flexible and robust reconstructions for MWT.


\section{Physics of Microwave Tomography}
\label{sec:physics}
The forward scattering problem of MWT establishes the mathematical relationship between the medium’s contrast and external measurements. For clarity, we consider the 2D transverse-magnetic (TM) case. Following the spectral method formulation \cite{burgel2017jcp}, we efficiently compute the forward operator and its derivatives. As illustrated in Fig.~\ref{fig:title_fig}(a), dielectric scatterers are placed within a square domain of interest (DOI) $\Omega_\text{D}\subset\mathbb{R}^2$, centered at the origin with side length $2d$. The computational domain (CD) is defined as $\Omega_{\text{C}}=[-2\sqrt{2}d, 2\sqrt{2}d)^2$, encompassing all computations. We denote by $B(r)$ the open ball centered at the origin with radius $r$, $H_l^{(1)}$ the Hankel function of the first kind and order $l$, $J_l$ the cylindrical Bessel function of order $l$, and $x$ the spatial coordinate. The Lebesgue and Sobolev spaces are denoted by $L^\alpha(\Omega_\text{D})$ and $W^{2,\beta}(\Omega_\text{D})$, respectively.

We consider a time-harmonic incident wave $u^\text{i}:\mathbb{R}^2\to\mathbb{C}$ with time dependence $\exp(-i\omega t)$, where $\omega > 0$ is the angular frequency. The medium’s contrast is defined as $\chi(x) = \epsilon_\text{r}(x) - 1$, where $\epsilon_\text{r}$ is the relative permittivity. The scattered field $u^\text{s}\in L^2(\Omega_\text{C})$ generated by $u^\text{i}$ satisfies the Lippmann-Schwinger (LS) equation:
\begin{equation}\label{eq:LS}
    u^\text{s}-\mathcal{V}_\text{C}\left(\chi\cdot u^\text{s}\right)=\mathcal{V}_\text{C}\left(\chi\cdot u^\text{i}\right),
\end{equation}
where $\mathcal{V}_\text{C}:L^2(\Omega_\text{C})\to L^2(\Omega_\text{C})$ is the periodized integral operator:
\begin{equation}
    \left(\mathcal{V}_\text{C}f\right)(x)=\int_{\Omega_\text{C}}\Phi_\text{C}(x-y)f(y)dy,\quad x\in\Omega_\text{C},
\end{equation}
with the periodized kernel $\Phi_\text{C}(x)=k^2i/4 \cdot H_0^{(1)}\left(k|x|\right),\ x\in B\left(2\sqrt{2}d\right)$ and vanishes when $x\in\Omega_\text{C}\backslash B\left(2\sqrt{2}d\right)$. The wavenumber $k$ is defined as $\omega\sqrt{\epsilon_0\mu_0}$, where $\epsilon_0$ and $\mu_0$ is the permittivity and the permeability of the free space respectively.

Since the contrast $\chi$ is supported in $\Omega_\text{D}$, Eq.~\eqref{eq:LS} requires only the incident wave in $\Omega_\text{D}$ to compute the scattered field in $\Omega_\text{D}$. This mapping is denoted by $T_\chi: L^\alpha(\Omega_\text{D}) \to W^{2,\beta}(\Omega_\text{D})$. The incident wave is modeled using point source single-layer potentials $\mathcal{S}_{\Gamma_\text{i}}: L^2(\Gamma_\text{i}) \to L^2(\Omega_\text{D})$ on the transmitter manifold $\Gamma_\text{i}$:
\begin{equation}
    \left(\mathcal{S}_{\Gamma_\text{i}}g\right)(x)=\frac{i}{4}\int_{\Gamma_\text{i}}H_0^{(1)}\left(k|x-y|\right)g(y)ds(y),\quad x\in\Omega_\text{D}.
\end{equation}
The near-field measurements are obtained via the operator $\mathcal{V}_\text{D}: L^2(\Omega_\text{D}) \to L^2(\Gamma_\text{s})$:
\begin{equation}
    \left(\mathcal{V}_\text{D}J\right)(x)=\int_{\Omega_\text{D}}\Phi_\text{C}\left(x-y\right)J(y)dy,\quad x\in\Gamma_\text{s},
\end{equation}
where $\Gamma_\text{s}$ is the receiver manifold and $J$ is the contrast source. The full forward operator $\mathcal{F}$, mapping the contrast $\chi\in L^p_{\text{Im}\ge 0}(\Omega_\text{D})$ to the corresponding measurements operator, is then given by $\mathcal{F}(\chi) = \mathcal{V}_\text{D} (\chi \cdot) T_\chi \mathcal{S}_{\Gamma_\text{i}}$, where $L^p_{\text{Im}\ge 0}(\Omega_\text{D})$ denotes the Lebesgue space with functions having non-negative imaginary part.

Given the contrast field $\chi$ and the direction $h$, the operator $\mathcal{F}$ is Fr\'{e}chet differentiable with 
\begin{equation}
    \mathcal{F}'(\chi)[h]=\mathcal{V}_\text{D}\left[I+\left(\chi\cdot\right)T_q\mathcal{V}_\text{C}\right]\left(h\cdot\right)T_q\mathcal{S}_{\Gamma_\text{i}},
\end{equation}
where $I$ is the identity. 

For numerical implementation, we discretize the CD using a mesh of $N^2$ regularly spaced collocation points with step size $h_N = 4\sqrt{2}d / N$. The scattered field and contrast in the DOI are represented by complex vectors $\underline{u}^\text{s}, \underline{\chi} \in \mathbb{C}^{N_\text{D}}$, where $N_\text{D} = \lfloor N / (2\sqrt{2}) \rfloor^2$. The discretized integral operator $\mathcal{V}_\text{C}$ restricted to the DOI is defined as $\mathcal{V}_{N_\text{D}}: \mathbb{C}^{N_\text{D}} \to \mathbb{C}^{N_\text{D}}$, given by $\mathcal{R}_\text{C} \text{FFT}^{-1} (\widehat{\Phi}_\text{C} \odot \cdot) \text{FFT} \mathcal{E}_\text{D}$, where $\text{FFT}$ is the 2D fast Fourier transform on $\mathbb{C}^{N \times N}$, $\mathcal{E}_\text{D}$ zero-extends grid points in $\Omega_\text{D}$ to $\mathbb{C}^{N \times N}$, and $\mathcal{R}_\text{C}$ restricts the matrix to $\Omega_\text{D}$. The kernel $\widehat{\Phi}_\text{C}$ is:

\begin{equation}
    \widehat{\Phi}_\text{C}(j)=
    \begin{cases}
        \begin{aligned}
            &\frac{\kappa^2}{\left(\pi|j|\right)^2-\kappa^2}\left(1+\frac{i\pi}{2}\left[\pi|j|J_1\left(\pi|j|\right)H_0^{(1)}(\kappa)\right.\right. \\ 
            &\quad\left.\left. -\kappa J_0\left(\pi|j|\right)H_1^{(1)}(\kappa)\right]\right),\  \text{if}\ \pi|j|\ne\kappa
        \end{aligned} \\
        \frac{i\pi\kappa^2}{4}\left[J_1(\kappa)H_1^{(1)}(\kappa)+J_0(\kappa)H_0^{(1)}(\kappa)\right],\ \text{else}
    \end{cases}
\end{equation}
where $\kappa = 2\sqrt{2}d \cdot k$ and $j$ represents integer lattice point. Using the Fourier basis in the spectral method, the discretized LS equation is:
\begin{equation}
    \underline{u}^\text{s}_{N_\text{D}}-\mathcal{V}_{N_\text{D}}\left(\underline{\chi}\odot\underline{u}^\text{s}_{N_\text{D}}\right)=\mathcal{V}_{N_\text{D}}\left(\underline{\chi}\odot\underline{u}^\text{i}_{N_\text{D}}\right),
\end{equation}
which can be tackled by GMRES algorithm in an iterative manner. The above process forms the discretization of the solution opertor $T_\chi$, denoted by $T_{\underline{\chi}}$. Also we denote by $\underline{\mathcal{V}}_\text{D}$, $\underline{\mathcal{S}}_{\Gamma_\text{i}}$ the standard discretization of the integral operator $\mathcal{V}_\text{D}$, $\mathcal{S}_{\Gamma_\text{i}}$ respectively.

The derivative of the discretized forward operator is:
\begin{equation}
    \mathcal{F}'\left(\underline{\chi}\right)[\underline{h}]=A_{N_\text{s},N_\text{D}}\left(\underline{h}\odot\right)B_{N_\text{D},N_\text{i}},
    \label{eq:f_prime}
\end{equation}
which is a matrix in $\mathbb{C}^{N_\text{s}\times N_\text{i}}$, where $N_\text{s}, N_\text{i}$ are the number of the receivers and the transmitters respectively, and $A_{N_\text{s},N_\text{D}}=\underline{\mathcal{V}}_\text{D}\left(I+\left(\underline{\chi}\odot\right)T_{\underline{\chi}}\mathcal{V}_{N_\text{D}}\right)$, $B_{N_\text{D},N_\text{i}}=T_{\underline{\chi}}\underline{\mathcal{S}}_{\Gamma_\text{i}}$. The corresponding adjoint of the derivative is given by
a matrix in $\mathbb{C}^{N_\text{s} \times N_\text{i}}$, where $N_\text{s}$ and $N_\text{i}$ are the number of receivers and transmitters, respectively, and $A_{N_\text{s}, N_\text{D}} = \underline{\mathcal{V}}_\text{D} (I + (\underline{\chi} \odot \cdot) T_{\underline{\chi}} \mathcal{V}_{N_\text{D}})$, $B_{N_\text{D}, N_\text{i}} = T_{\underline{\chi}} \underline{\mathcal{S}}_{\Gamma_\text{i}}$. The adjoint of the derivative is:
\begin{equation}
    \left[\mathcal{F}^\prime\left(\underline{\chi}\right)\right]^* \left[\underline{H}\right]=\sum_{j=1}^{N_\text{s}}\sum_{l=1}^{N_\text{i}}\underline{H}_{j,l}\overline{A_{N_\text{s},N_\text{D}}(j,\cdot)}\overline{B_{N_\text{D},N_\text{i}}(\cdot,l)}.
    \label{eq:adjoint}
\end{equation}
The contrast is optimized using gradient-based methods with this adjoint.

\begin{figure*}
  \centering
  \includegraphics[width=1.\textwidth]{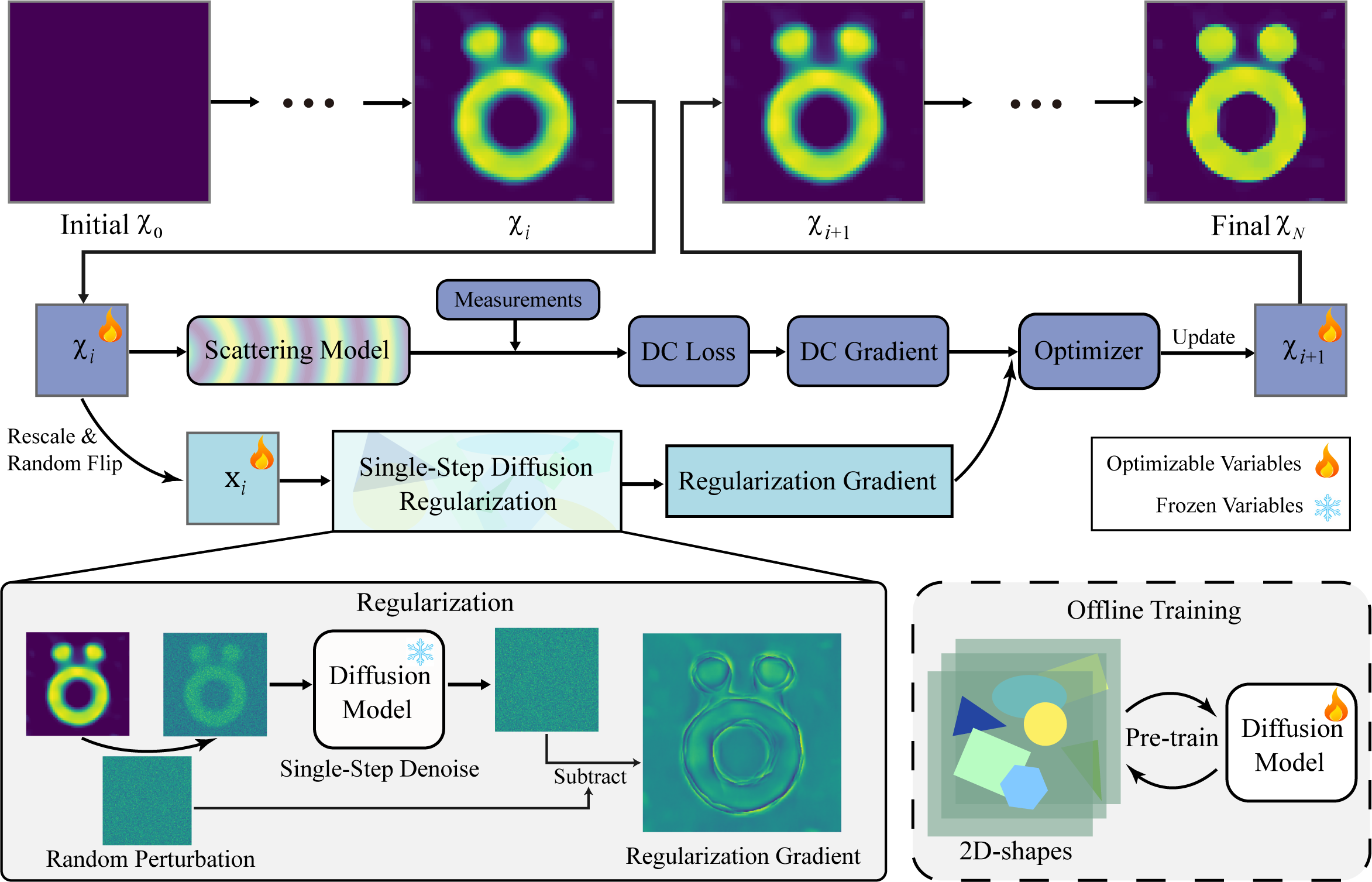}
  \caption{Overview of the proposed reconstruction framework. The method solves a DC-driven optimization problem with SSD-based regularization. At each iteration, the contrast distribution $\chi$ is processed through two branches: the forward modeling branch computes the DC gradient via the physical model, while the regularization branch uses a pre-trained DM to compute the regularization gradient. The combined gradient updates $\chi$, and the process repeats until convergence. }
  \label{fig:structure}
\end{figure*}
\section{Method}
\label{sec:method}
This section presents the proposed reconstruction approach for MWT.
First, we formulate the MWT inverse problem as an optimization procedure with Fr\'{e}chet-differentiable forward operators. This differentiable formulation enables gradient-based optimization while ensuring physical consistency (Sec.~\ref{sec:method_1}).
Second, we introduce a DM-based regularization to incorporate learned priors. The random flipping strategy is applied during regularization to mitigate dataset bias (Sec.~\ref{sec:method_2}).
A schematic overview is shown in Fig.~\ref{fig:structure}.

\subsection{Data-Consistency Guided Optimization}
\label{sec:method_1}
MWT reconstruction is usually formulated as an optimization problem with the objective:
\begin{equation}
  \min_{\underline{\chi}} \left(\frac{1}{2} \| \mathcal{F}(\underline{\chi})  - {u}^{\mathrm{sct}}_\mathrm{meas} \|_2^2 + \lambda R(\underline{\chi}) \right),
  \label{eq:optim_obj}
\end{equation}
where $u^\mathrm{sct}_\mathrm{meas} \in \mathbb{C}^{N_\text{s} \times N_\text{i}}$ represents the measured scattered field, $\mathcal{F}(\underline{\chi})$ is the forward operator mapping the discrete contrast $\underline{\chi}$ to predicted measurements, $R(\underline{\chi})$ is the regularization term, and $\lambda$ is its weighting factor. The data-consistency term is defined as $\mathcal{L}_\mathrm{DC} = \frac{1}{2} \|\mathcal{F}(\underline{\chi}) - u^\mathrm{sct}_\mathrm{meas}\|_2^2$.
To optimize this objective, we employ gradient-based methods, leveraging the gradient of the data-consistency term and the regularization. The gradient of $\mathcal{L}_\mathrm{DC}$ is computed using the adjoint of the Fr\'{e}chet derivative of the forward operator, as derived in Eq.~\eqref{eq:adjoint}:
\begin{align}
  \nabla_{\underline{\chi}}\mathcal{L}_{DC} &= [\mathcal{F}^\prime (\underline{\chi})]^* [\underline{H}] = A^* \cdot \underline{H} \cdot B^*, \\
  \underline{H} &= \mathcal{F}(\underline{\chi}) - {u}^{\mathrm{sct}}_\mathrm{meas}, \quad \underline{H} \in \mathbb{C}^{N_\text{s} \times N_\text{i}},
  \label{eq:residual}
\end{align}
where $A \in \mathbb{C}^{N_\text{s} \times N_\text{D}}$ and $B \in \mathbb{C}^{N_\text{D} \times N_\text{i}}$ are Jacobian matrices, and $A^*$ and $B^*$ are their conjugate transposes. The residual $H$ represents the difference between predicted and measured scattered fields. The gradient $\nabla_{\underline{\chi}} \mathcal{L}_\mathrm{DC}$ is combined with the regularization gradient $\nabla_{\underline{\chi}} R(\underline{\chi})$ and used in an optimizer, such as Adam, to update the contrast $\underline{\chi}$.

\subsection{Single-Step Diffusion Regularization}
\label{sec:method_2}
To address the ill-posedness of MWT and introduce learned prior, we propose a regularization term based on a pretrained DM, termed SSD-Reg. The dielectric contrast $\underline{\chi}$ is first normalized to represent an image $\mathbf{x}_0$ at the initial time step of the diffusion process. To mitigate bias accumulation and enhance robustness, $\mathbf{x}_0$ is subjected to random data augmentations, including horizontal and vertical flips.
The SSD-Reg is defined as:
\begin{equation}
  R(\mathbf{x}_0) = \lambda_t \ \text{sg}[\xi_\phi(\mathbf{x}_t, t) - \xi_t]^\mathsf{T}\cdot \mathbf{x}_0,\quad\lambda_t = \lambda / \text{SNR}_t,
  \label{eq:ssdreg}
\end{equation}
where $\text{sg}[\cdot]$ denotes the stop-gradient operator, $\xi_\phi(\mathbf{x}_t, t)$ is the noise prediction from a pretrained DM at time step $t$, $\xi_t$ is the random noise added at $t$, and $\text{SNR}_t = \sqrt{1 - \sigma_t^2} / \sigma_t$ is the signal-to-noise ratio, with $\sigma_t$ being the noise schedule parameter \cite{reddiff}.
With stop-gradient applied, the gradient of $R(\mathbf{x}_0)$ becomes $\lambda_t \ \text{sg}[\xi_\phi(\mathbf{x}_t, t) - \xi_t]^\top$, which corresponds to a single-step denoising process regularizing $\mathbf{x}_0$.

The pretrained DM, trained on a dataset of complex shape images, learns to capture the statistical distribution of plausible reconstructions, enabling robust regularization without requiring multiple denoising steps. This single-step approach reduces computational overhead while maintaining the benefits of diffusion-based priors.

The total loss function combines the DC and SSD terms:
\begin{align}
  \mathcal{L} &= \mathcal{L}_{DC} + \mathcal{L}_{SSD} \\
              &= \frac{1}{2} \| \mathcal{F}(\underline{\chi}) - {u}^{\mathrm{sct}}_\mathrm{meas} \|_2^2 + \lambda_t\ \text{sg}[\xi_\phi(\mathbf{x}_t, t) - \xi_t]^\mathsf{T} \cdot \mathbf{x}_0,
  \label{eq:total_loss}
\end{align}
where $\mathcal{L}_{DC}$ denotes the data consistency loss and $\mathcal{L}_{SSD}$ represents the regularization loss. For gradient computation, the forward-problem gradient is manually calculated, while all other gradient components are derived via automatic differentiation. The resulting composite gradient is then used by the Adam optimizer to optimize the contrast. This design maintains computational accuracy in critical components while fully leveraging the automatic differentiation capabilities of modern deep learning frameworks.

\subsection{Implementation Details}
We implemented the forward solver in Python based on the MATLAB code from \cite{burgel2017jcp}. The SSD framework was primarily developed using PyTorch, with the exception of the forward solver and Jacobian calculations, which were computed using SciPy. For the diffusion prior, we employed the pre-trained model from \cite{tong2024diffinr}, keeping its parameters fixed during reconstruction without fine-tuning.

To improve computational efficiency, the Jacobian $B$ was updated at each iteration while Jacobian $A$ was updated every five steps. For SSD, the regularization weight $\lambda=0.5$, the learning rate $lr=0.1$, the diffusion time step $t$ was linearly annealed from 500 to 1, and the maximum iterations $N_\text{max}=500$. All experiments were conducted on a workstation equipped with an Intel Core i7-12700K CPU and an NVIDIA GeForce RTX 4070 GPU.
The complete reconstruction algorithm is outlined in Algorithm~\ref{Alg.SSD}. 

\begin{algorithm}
\caption{SSD-Reg for Stable MWT Reconstruction}
\label{Alg.SSD}
\begin{algorithmic}
\State \textbf{Input:} Measured scattered field ${u}^{\mathrm{sct}}_\mathrm{meas}$, MWT forward model $\mathcal{F}$, regularization weight $\lambda_t$, maximum iterations $N_\text{max}$, pretrained DM $\xi_\phi$, diffusion schedule $\alpha_t$
\State \textbf{Initialize:} $\underline{\chi}$

\For{$i = 1$ \textbf{to} $N$}
    \State Compute forward model ${u}^{\mathrm{sct}}_\mathrm{RX} =\mathcal{F}(\underline{\chi})$
    \State Residual $H = {u}^{\mathrm{sct}}_\mathrm{RX}   - {u}^{\mathrm{sct}}_\mathrm{meas}$
    \State Compute $\mathcal{L}_{DC} = \frac{1}{2} \| H \|_2^2$
    \If{$i > 200$ and change in $\mathcal{L}_{DC} < 0.001$}
    \State Terminate and return $\underline{\chi}$
    \EndIf
    \State Compute $B$
    \If{$i \bmod 5 == 0$}
        \State Compute $A$
    \EndIf
    \State Compute adjoint gradient: $\nabla_{\underline{\chi}}\mathcal{L}_{DC} = A^* \cdot \underline{H} \cdot B^*$
    \State Get a time step $t$ and random noise $\xi_t$
    \State Normalize and random flip $\underline{\chi} \rightarrow \mathbf{x}_0$
    \State Perturb image: $\mathbf{x}_t = \sqrt{\alpha_t} \mathbf{x}_0 + \sqrt{1 - \alpha_t} \xi_t$
    \State Compute $\mathcal{L}_{SSD} = \lambda_t \ \text{sg}[\xi_\phi(\mathbf{x}_t, t) - \xi_t]^\mathsf{T} \cdot \mathbf{x}_0$
    \State Auto-gradient $\nabla_{\underline{\chi}}\mathcal{L}_{SSD}$
    \State Compute total gradient $\nabla_{\underline{\chi}}\mathcal{L}=\nabla_{\underline{\chi}}\mathcal{L}_{DC}+\nabla_{\underline{\chi}}\mathcal{L}_{SSD}$
    \State Update $\underline{\chi}$ using Adam($\nabla_{\underline{\chi}} \mathcal{L}$)
\EndFor
\end{algorithmic}
\end{algorithm}

\section{Experiments}
\label{sec:exp}
This section evaluates the performance of the proposed SSD-based reconstruction framework through extensive comparative experiments on both synthetic and real-world datasets. We benchmark our method against five approaches to assess its accuracy, robustness, and computational efficiency.

\subsection{Experimental Setup}
\subsubsection{Datasets and Test Cases}
We pretrained the DM on a dataset of 50,000 synthetic images comprising up to five overlapping polygons and closed B\'{e}zier curves with varied geometries and intensities \cite{tong2024diffinr}. Due to its complexity, this dataset is not suitable for direct permittivity reconstruction, which motivates the creation of application-specific datasets:
\begin{itemize}
\item \textbf{Simulated Data}:
\textbf{Austria Data}: A standard EISP test case used to evaluate geometric reconstruction \cite{zhou2022deep}.
\textbf{2D Shape Data}: Ten cases of basic shapes (triangles, ellipses, polygons) in overlapping or non-overlapping configurations, with relative permittivity of 1.3--2.5.
\textbf{MNIST Data}: Ten MNIST handwritten digit images with permittivity settings matching the 2D shapes.
All simulations use a 2 m $\times$ 2 m region, with 16 transmitters and 32 receivers on a 3 m-radius circular boundary, operating at 400 MHz \cite{wei2018bps, luo2024inr_eisp}.
\item \textbf{High-Contrast Human Breast Dataset}: A 20 cm $\times$ 20 cm DOI with a coupling medium of relative permittivity 10, featuring tissues with high permittivity contrast. Uses 16 transmitters and 32 receivers on a 0.5 m-radius circle, with frequencies of 1.0, 1.4, 1.8, and 2.2 GHz.
\item \textbf{Fresnel Dataset}: Real-world scattering data from dielectric materials (nylon and foam) measured in a microwave anechoic chamber \cite{geffrin2005fresnel2}. Configurations include 8 to 18 transmitters and 241 receivers. For all experiments, we utilize 5 GHz TM-mode data.
\end{itemize}

To approximate real-world measurement scenarios, we added 5\% Gaussian noise to all synthetic datasets, with the exception of our noise-level robustness evaluation in Sec. \ref{subsec:noiserobust} where varying noise conditions were examined.

\subsubsection{Comparing Methods}
We compare the proposed SSD method against five baselines:
{\bf BP}: The backpropagation (BP) algorithm, a classical analytical non-iterative reconstruction method \cite{belkebir2005bp}.
{\bf PDA}: A Primal-Dual Algorithm (PDA) with sparsity regularization \cite{burgel2017jcp}, a traditional iterative method.
{\bf BPS}: A supervised CNN-based post-processing method using BP results as input \cite{wei2018bps}. 
{\bf INR+TV}: A recent SOTA method leveraging INR with TV regularization \cite{luo2024inr_eisp}.
{\bf DPS}: Diffusion Posterior Sampling (DPS) method for general inverse problems \cite{chung2023dps}, we adapt this method to MWT reconstruction.

All methods, except BPS, operate in an unsupervised manner and are individually optimized for performance.

\subsubsection{Evaluation Metrics}
\label{sec:metrics}
We assess reconstruction quality using Peak Signal-to-Noise Ratio (PSNR), Structural Similarity Index (SSIM) \cite{ssim_2004tip}, and Learned Perceptual Image Patch Similarity (LPIPS) \cite{zhang2018lpips}. To provide a unified measure of performance, we define a composite score $\mathcal{S}$, as show in Fig.\ref{fig:title_fig}(b), integrating the normalized forms of these metrics.

For each method $i$ among $N_\text{methods}$ compared, we first normalize the metrics as follows:
\begin{align}
    \mathrm{SSIM}_{\mathrm{norm}, i} &= \frac{\mathrm{SSIM}_i - \min \{\mathrm{SSIM}\}}{\max \{\mathrm{SSIM}\} - \min \{\mathrm{SSIM}\}}, \\
    \mathrm{PSNR}_{\mathrm{norm}, i} &= \frac{\mathrm{PSNR}_i - \min \{\mathrm{PSNR}\}}{\max \{\mathrm{PSNR}\} - \min \{\mathrm{PSNR}\}}, \\
    \mathrm{LPIPS}_{\mathrm{norm}, i} &=1- \frac{\mathrm{LPIPS}_i - \min \{\mathrm{LPIPS}\}}{\max \{\mathrm{LPIPS}\} - \min \{\mathrm{LPIPS}\}},
\end{align}
where $\text{metric}_i$ represents the average value of the metric for the $i$-th method. 
The composite score $\mathcal{S_i}$ is then computed as the arithmetic mean of the normalized metrics:
\begin{equation}
    \label{eq:score}
    \mathcal{S_i} = \frac{1}{3}\left(\mathrm{SSIM}_{\mathrm{norm}, i} + \mathrm{PSNR}_{\mathrm{norm}, i} + \mathrm{LPIPS}_{\mathrm{norm}, i}\right)
\end{equation}

\begin{figure*}
  \centering
  \includegraphics[width=1.\textwidth]{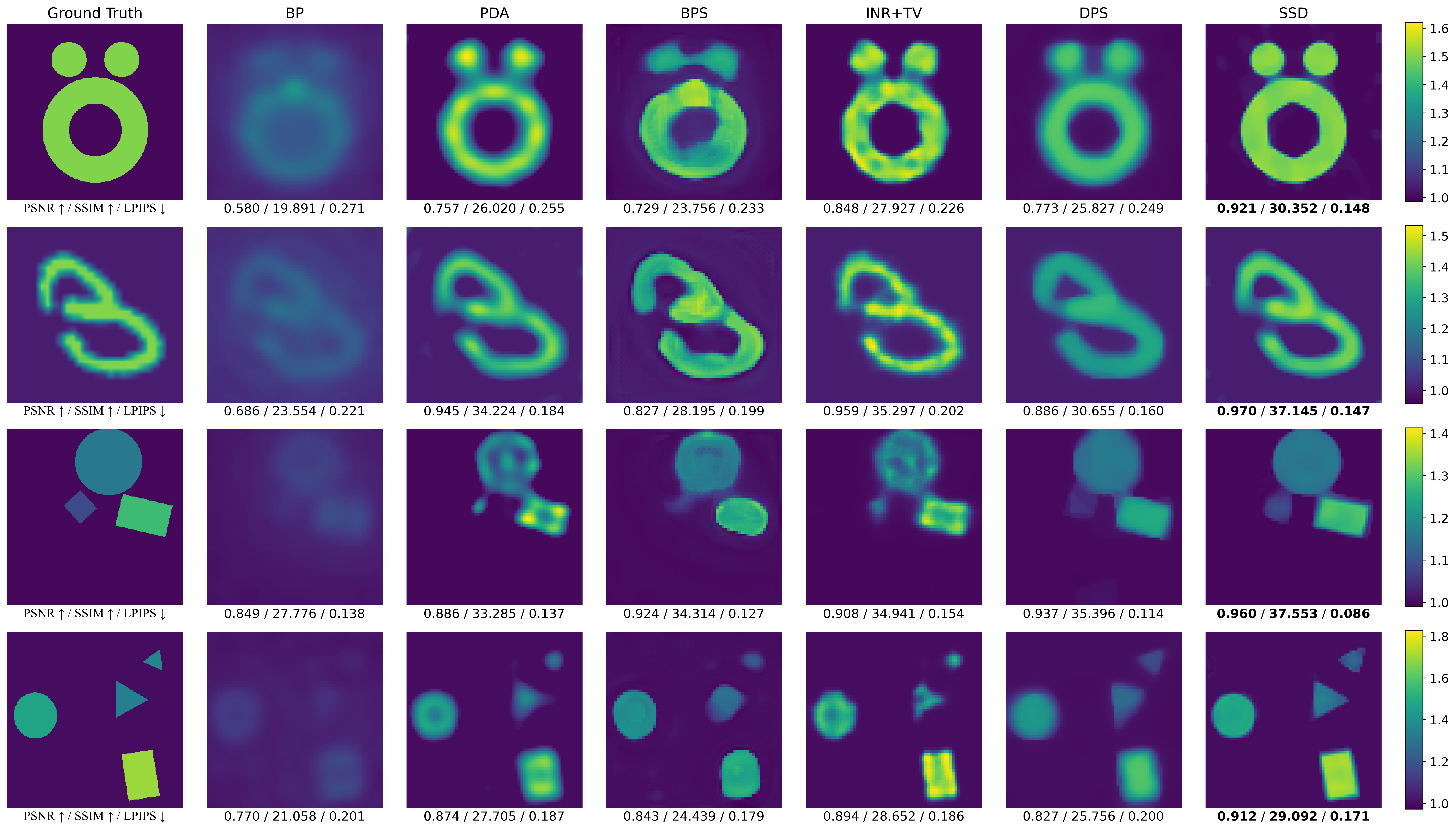}
  \caption{Reconstruction results on simulated data, with quantitative metrics (PSNR/SSIM/LPIPS) shown below each image; bold values indicate the best performance. Colorbars denote relative permittivity. From top to bottom: Austria test case, one MNIST sample, and two samples from the 2D shapes dataset.}
  \label{fig:simu_exp1}
\end{figure*}

\subsection{Comparison on Simulation Data}

As shown in Fig.~\ref{fig:simu_exp1}, BP produces low-quality reconstructions characterized by blurred shapes and significant numerical inaccuracies. In contrast,  PDA yields substantially improved reconstructions but introduces speckle-like artifacts in regions with homogeneous permittivity, particularly evident in the first and third rows.

The supervised post-processing approach BPS performs well on in-distribution (ID) samples but generalizes poorly to out-of-distribution (OOD) cases, such as the Austria test case, where it inherits geometric distortions from the initial BP reconstructions.

The INR+TV method outperforms traditional approaches, but similar to PDA, its reliance solely on data fidelity and TV regularization leads to periodic artifacts in the reconstructed images.

Incorporating diffusion-based priors, DPS and SSD significantly suppress these artifacts and generate shapes more consistent with the GT. However, DPS often underperforms INR+TV in quantitative metrics, suggesting that direct constraint via diffusion sampling can be suboptimal. Specifically, the use of DC loss gradients to steer the sampling trajectory destabilizes convergence due to the ill-posed nature of MWT. As a result, DPS reconstructions exhibit substantial permittivity deviations and unreliable shape boundaries.

In contrast, SSD achieves accurate shape recovery while preserving numerical fidelity, consistently outperforming other methods across quantitative metrics. This improvement stems from SSD’s integration of shape priors as a regularization term, which augments rather than interferes with the DC-driven optimization process.

\begin{figure*}
  \centering
  \includegraphics[width=0.95\textwidth]{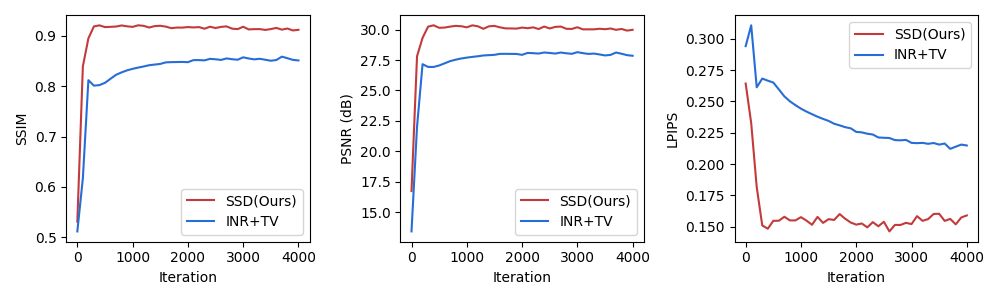}
  \caption{Reconstruction metrics comparison between SSD and INR+TV during iterative optimization on the Austria test case.}
  \label{fig:iteration}
\end{figure*}

We further compare SSD and INR+TV in terms of convergence behavior and reconstruction speed. As shown in Fig.~\ref{fig:iteration}, SSD converges within approximately 200 iterations, whereas INR+TV requires over 2000 iterations to converge. The faster convergence of SSD is attributed to both its regularization strategy and the forward model formulation. SSD-Reg mitigates the inherent ill-posedness of the inverse problem, and the incorporation of a differentiable forward model imposes a stronger physical constraint compared to the CSI-based approach employed in INR+TV. This combination enables efficient gradient descent and accelerates convergence.

We also evaluated the average reconstruction time of unsupervised methods across all simulation cases. The mean times per case were 45.22 s for SSD, 58.64 s for DPS, and 408.72 s for INR+TV. SSD achieved a 9$\times$ speedup over INR+TV, demonstrating clear computational advantages.
This efficiency gain arises from several factors: SSD avoids the parameter-intensive implicit representation used in INR+TV, reducing optimization complexity and allowing for larger learning rates; the accurate Jacobian matrices computed from the forward model provide well-informed descent directions that accelerate convergence; and the single-step inference mechanism of SSD-Reg enables efficient integration of learned priors.
In practice, SSD typically terminates after 200 to 300 iterations with stopping criterion defined by changes in DC loss. This criterion mitigates overfitting to measurement noise and forward model errors while maintaining stable performance. Fig.~\ref{fig:iteration} show that further optimization beyond this point yields minimal improvements and may even degrade performance.

\begin{figure}
  \centering
  \includegraphics[width=0.5\textwidth]{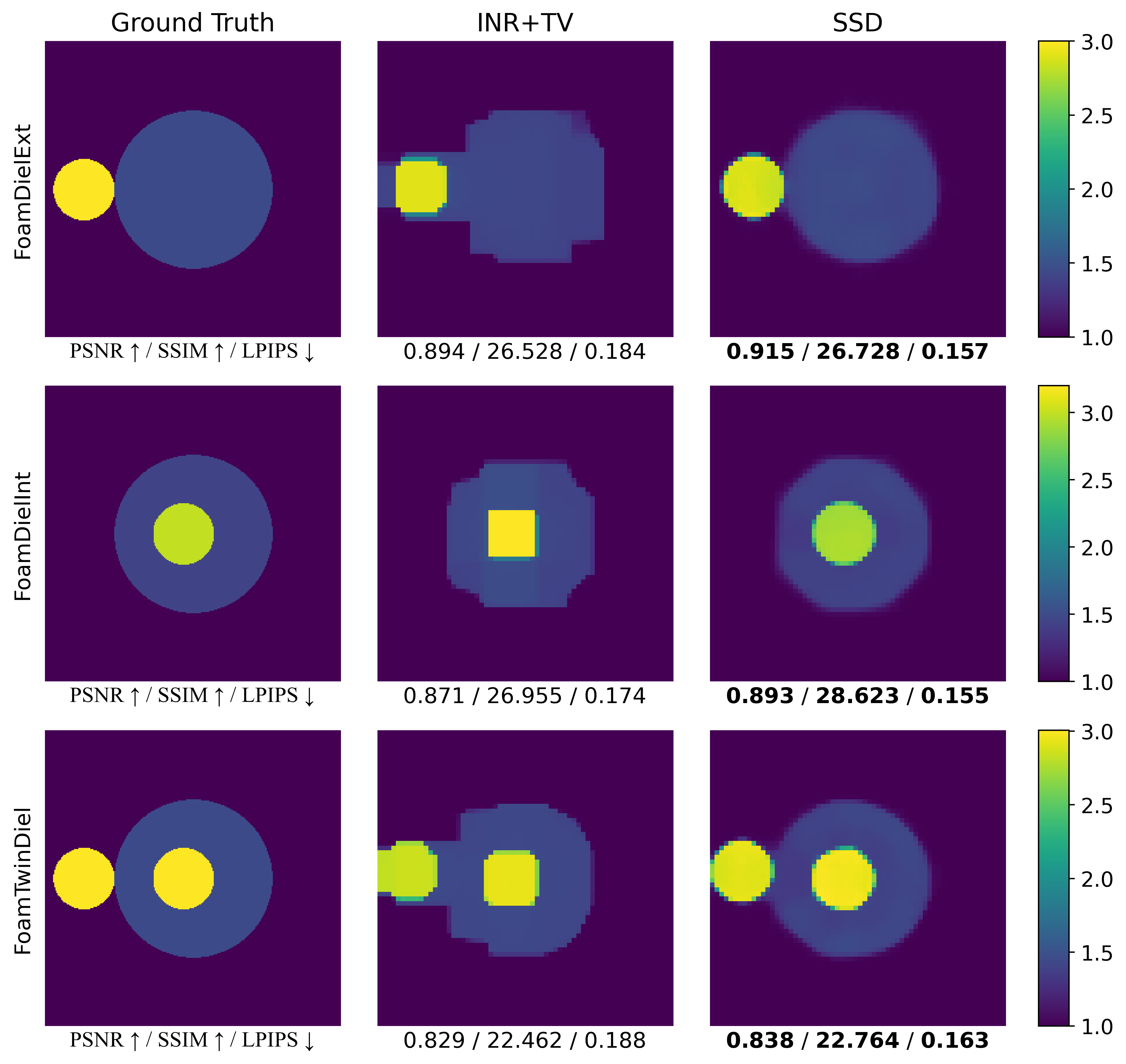}
  \caption{Reconstruction of INR+TV and SSD on real-world data, with metrics displayed beneath each result, bold values indicate the best performance. Colorbars indicate relative permittivity.}
  \label{fig:real_world_exp}
\end{figure}
\subsection{Comparison on Real-World Data}
We conducted three reconstructions using real-world measurement data, as shown in Fig.~\ref{fig:real_world_exp}. Our method demonstrates superior shape reconstruction accuracy compared to the previous SOTA approach, INR+TV. The INR+TV method suffers from over-reliance on TV regularization, leading to blocky staircasing artifacts in the reconstructed shapes. In contrast, SSD effectively avoids such artifacts while maintaining reconstruction fidelity.

In the FoamTwinDiel case, experimental setup errors cause the connecting line between the two nylon rods to deviate slightly from perpendicular alignment with the DOI boundary. This misalignment is observable across all reconstruction methods, including those reported in \cite{luo2024inr_eisp}. Consequently, quantitative metrics based on GT images may not fully reflect reconstruction quality. Visual assessment provides more intuitive performance comparison, clearly demonstrating SSD's advantages: it achieves accurate relative permittivity reconstruction and high-fidelity shape recovery in real-world scenarios.

\section{Discussions}
\label{sec:discussions}
This section evaluates the noise robustness and preprocessing strategies of the proposed SSD framework, examines its performance in high-contrast scenarios relevant to clinical applications, and discusses its limitations and potential directions for future research.

\begin{figure}
  \centering
  \includegraphics[width=0.7\textwidth]{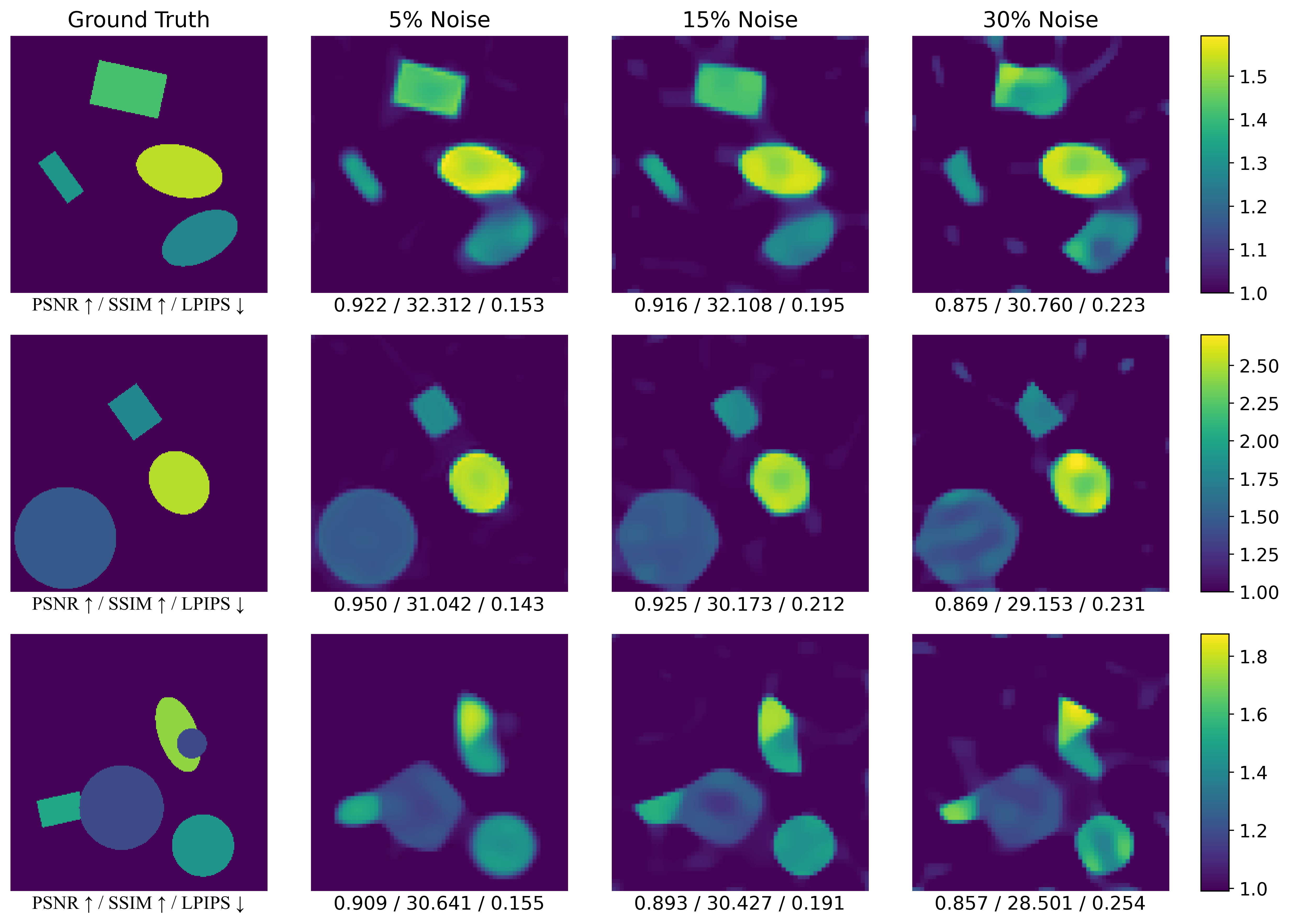}
  \caption{Three representative reconstruction results of the SSD method under different noise levels (5\%, 15\%, and 30\%) on simulated data with metrics displayed beneath each result. The Colorbar indicate relative permittivity.}
  \label{fig:noise_robustness}
\end{figure}

\subsection{Noise Robustness}
\label{subsec:noiserobust}
To assess the robustness of SSD under high-noise conditions, we performed additional reconstructions on the simulated shapes dataset with 15\% and 30\% additive Gaussian noise. Quantitative results are summarized in Table~\ref{table:noise}, and Fig.~\ref{fig:noise_robustness} visually demonstrates SSD-Reg’s ability to preserve structural features and permittivity values as noise increases.

\begin{table*}[ht]
    \centering
    \caption{Quantitative comparison of reconstruction performance under noise}
    \begin{tabular*}{.85\textwidth}{@{\extracolsep{\fill}} c|ccc|ccc|ccc }
        \hline
         & \multicolumn{3}{c|}{\textbf{5\% Noise}} & \multicolumn{3}{c|}{\textbf{15\% Noise}} & \multicolumn{3}{c}{\textbf{30\% Noise}} \\
        & \textbf{SSIM} & \textbf{PSNR} & \textbf{LPIPS} & \textbf{SSIM} & \textbf{PSNR} & \textbf{LPIPS} & \textbf{SSIM} & \textbf{PSNR} & \textbf{LPIPS} \\
        \hline
        
        SSD    & \textbf{0.933} & \textbf{32.47} & \textbf{0.134} & \textbf{0.928} & \textbf{32.43} & 0.181 & \textbf{0.882} & \textbf{30.48} & 0.236 \\

        DPS    & 0.870 & 28.90 & 0.170 & 0.869 & 28.86 & \textbf{0.171} & 0.864 & 28.74 & \textbf{0.173} \\

        INR+TV & 0.907 & 31.44 & 0.172 & 0.887 & 30.23 & 0.183 & 0.837 & 27.91 & 0.198 \\

        BPS    & 0.853 & 27.60 & 0.167 & 0.850 & 27.52 & 0.171 & 0.840 & 27.25 & 0.179 \\

        PDA    & 0.875 & 29.73 & 0.177 & 0.822 & 27.47 & 0.246 & 0.750 & 24.92 & 0.295 \\
        
        BP     & 0.774 & 22.87 & 0.186 & 0.774 & 22.86 & 0.187 & 0.772 & 22.82 & 0.188 \\

    \end{tabular*}
    \label{table:noise}
\end{table*}

As shown in Table~\ref{table:noise}, SSD consistently achieves the highest SSIM and PSNR across all noise levels, outperforming iterative methods such as INR+TV and PDA, which exhibit significant degradation at higher noise due to their strong reliance on data fidelity. Even at 30\% noise, where the low signal-to-noise ratio affects all methods, SSD-Reg maintains superior reconstruction quality.

Compared to DPS, which uses the same DM, SSD-Reg offers better SSIM and PSNR, reflecting its stronger balance between data consistency and learned priors. While DPS benefits from noise robustness through strong generative priors, it tends to over-smooth fine details, particularly in low-noise scenarios where data-driven gradients are more informative. In contrast, SSD-Reg’s PnP regularization achieves more flexible integration, enabling superior performance across both low- and high-noise regimes.
Although SSD shows worse LPIPS scores at high noise levels due to minor speckle-like artifacts, SSIM and PSNR remain stable, which are more representative of structural and quantitative accuracy in MWT. DPS’s smoother textures yield better LPIPS but come at the expense of reconstruction fidelity, limiting its practical utility.

As demonstrated in Fig.\ref{fig:noise_robustness}, SSD effectively retains critical structural features and accurate permittivity estimation even under severe noise. Together with Table \ref{table:noise}, these results highlight that the observed LPIPS degradation mainly stems from subtle low-intensity artifacts, which do not significantly compromise clinical interpretability.

In conclusion, SSD demonstrates strong robustness to noise, delivering reliable and accurate reconstructions across a wide range of noise levels. This confirms its practical value for real-world MWT applications, where measurement conditions are often imperfect and variable.

\begin{figure}
  \centering
  \includegraphics[width=0.5\textwidth]{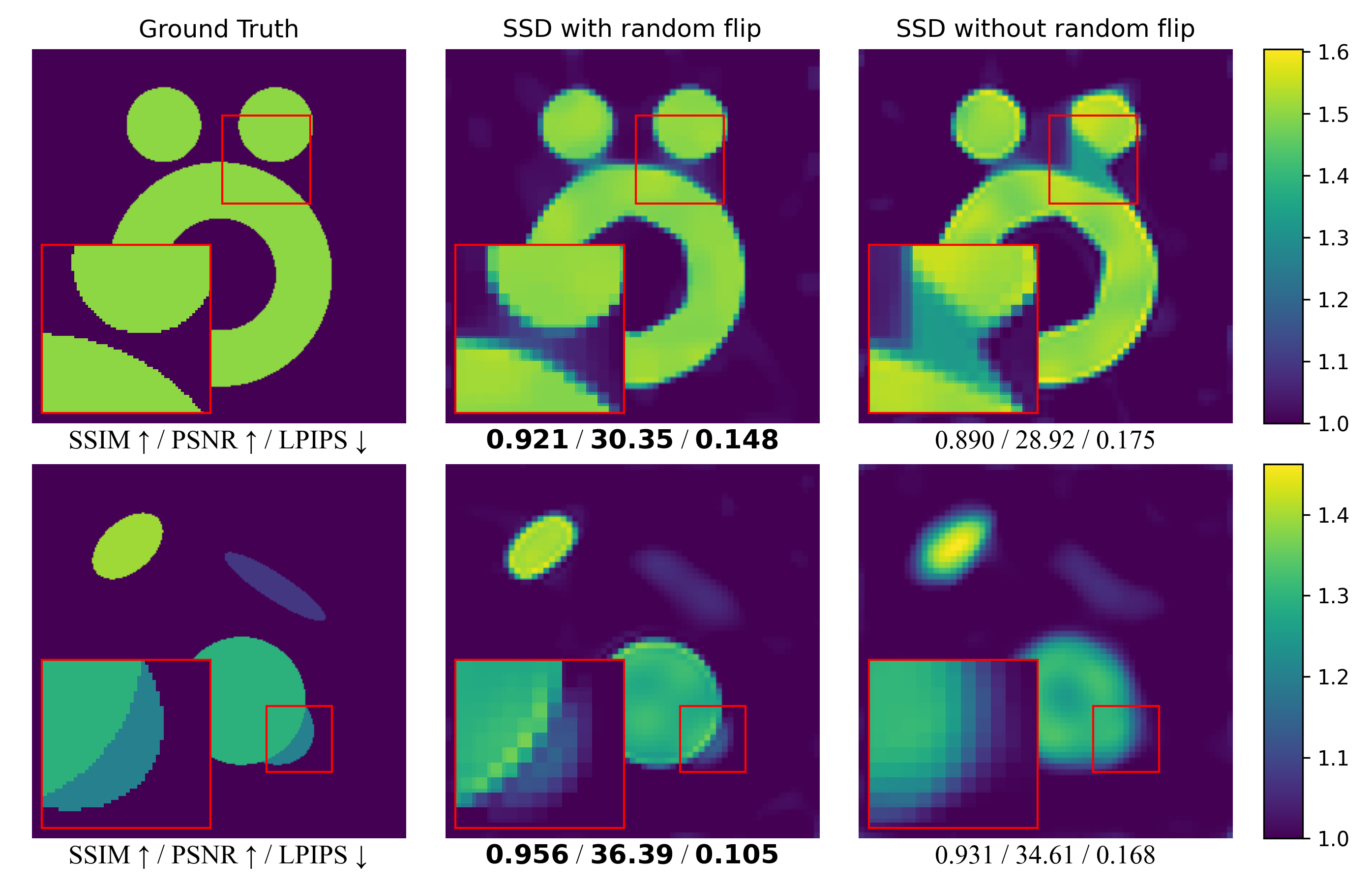}
  \caption{Ablation study comparing the SSD reconstruction performance with and without the random flipping mechanism.}
  \label{fig:ablation}
\end{figure}

\subsection{Ablation Study on Random Flipping}
\begin{table}[ht]
    \centering
    \caption{Ablation study on random flipping}
    \begin{tabular}{c|ccc}
        \hline
        & SSIM & PSNR & LPIPS \\
        \hline
        SSD with random flip    & \textbf{0.933} & \textbf{32.47} & \textbf{0.134} \\
        SSD without random flip & 0.919 & 31.63 & 0.168 \\
        \hline
    \end{tabular}
    \label{table:ablation}
\end{table}
Random flipping, the primary preprocessing strategy in SSD-Reg, mitigates dataset bias and enhances generalization. Ablation experiments, summarized in Table~\ref{table:ablation}, show that removing random flipping degrades SSIM, PSNR, and LPIPS. Fig.~\ref{fig:ablation} illustrates error accumulation without random flipping, such as geometric distortion in the Austria dataset (e.g., circular structures reconstructed as polygons) and deformed shapes in the 2D shapes dataset (e.g., incorrect fusion of occluded circles). These results confirm the critical role of random flipping in ensuring robust reconstruction.

\subsection{High contrast Reconstruction}
\begin{figure}
  \centering
  \includegraphics[width=0.5\textwidth]{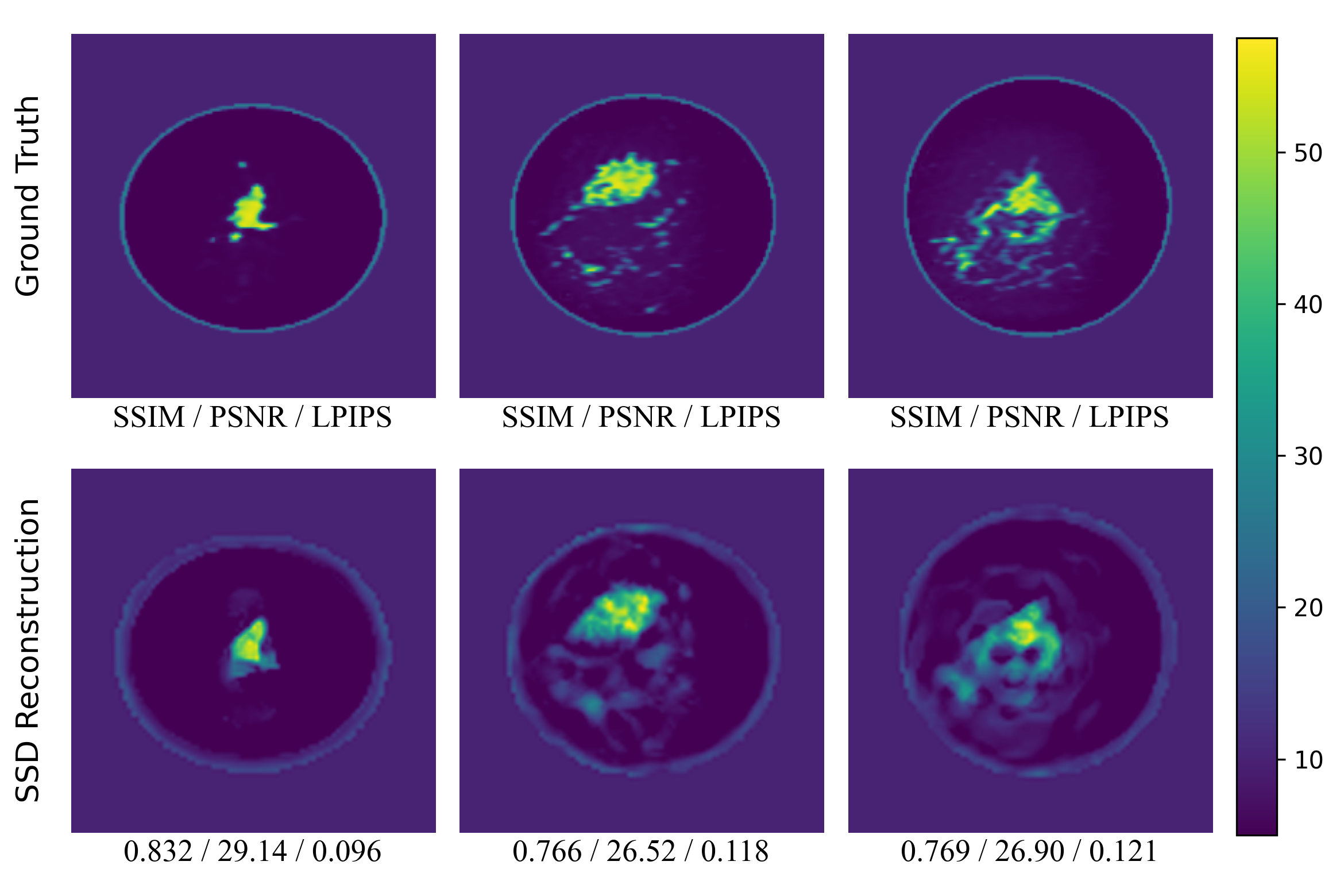}
  \caption{SSD reconstruction on breast phantom with metrics displayed beneath each result. The Colorbar indicate relative permittivity.}
  \label{fig:high_contrast}
\end{figure}

While MWT is often optimized for low-contrast scenarios \cite{xd_chen2018isp}, but high-contrast scenarios, such as those in breast imaging \cite{kwon2016mwi_brst_cnsr, mojabi2025mwi_3d_breast}, are critical for clinical applications. Tumor tissues typically exhibit relative permittivity more than five times higher than that of surrounding normal tissues. However, high-contrast scenarios pose significant challenges, requiring precise initial solutions and meticulous parameter optimization \cite{xd_chen2018isp}. As demonstrated in Fig.~\ref{fig:high_contrast}, using breast phantoms with contrasts exceeding 5 between abnormal and surrounding tissues, our proposed SSD method achieves robust reconstruction, accurately recovering both the morphology and relative permittivity of abnormal tissues.

Unlike INR-based methods \cite{luo2024inr_eisp}, which rely on output activation functions and manually tuned scaling coefficients, often leading to all-zero outputs or divergence in high-contrast cases, SSD adopts direct pixel-wise updates guided by accurate Jacobians. This design enables stable reconstruction even from an all-zero initialization.
The successful differentiation between low-permittivity normal tissues and high-permittivity abnormal tissues in breast phantom experiments underscores SSD’s robustness and clinical relevance. These results highlight its potential for real-world applications, such as early tumor detection and stroke localization.

\subsection{Limitations and Future Research Directions}
While SSD demonstrates strong and robust performance across various scenarios, certain limitations remain. Currently, the method primarily provides geometric regularization without incorporation of task-specific anatomical priors, which may reduce effectiveness in highly complex or heterogeneous anatomical contexts. To address this, future research will explore the integration of anatomically informed priors by leveraging latent diffusion models conditioned on multi-modal imaging data and semantic information. This approach aims to enable more precise, application-specific regularization, particularly benefiting clinically critical domains such as breast and brain imaging.

In addition, extending the framework to 3D microwave imaging is a natural next step, as the forward model is not restricted to two dimensions. SSD-Reg can take advantage of parallel GPU processing by applying regularization across 2D slices for efficient 3D reconstruction. We also plan to investigate more memory-efficient representations to support scalable, high-resolution 3D imaging.

\section{Conclusion}
\label{sec:conclusion}
In this work, we presented SSD-Reg, a learned regularization approach for microwave tomography that integrates shape priors from DMs without the need for paired training data. By embedding these priors into the reconstruction process via a PnP framework, SSD-Reg effectively mitigates the ill-posedness of the inverse problem. Leveraging a Fr\'{e}chet-differentiable forward operator, the proposed method SSD achieves improved stability, robustness, and faster convergence in iterative reconstruction. Extensive evaluations on both simulated and real-world datasets confirm that SSD enables reliable and efficient reconstruction, highlighting its strong potential for clinical applications.

\bibliographystyle{IEEEtran}
\bibliography{reference}
\end{document}